\documentclass[useAMS,usenatbib,12pt,epsfig,letter]{mn2e}
\usepackage[english]{babel}
\usepackage{amssymb}
\usepackage{graphicx}

\title
[Magnification and dust reddening]
{Measuring the galaxy-mass and galaxy-dust correlations through magnification and reddening} 
\author[M\'enard et al.]
{
\parbox[h]{\textwidth}{
Brice M\'enard$^{1}$, 
Ryan Scranton$^2$, 
Masataka Fukugita$^{3,4}$, 
Gordon Richards$^5$}
\vspace*{2pt} \\
\hspace{-.1cm}$^1$ Canadian Institute for Theoretical Astrophysics\\
\hspace{-.1cm}$^2$ University of California-Davis\\
\hspace{-.1cm}$^3$ Institute for Advanced Study\\
\hspace{-.1cm}$^4$ Tokyo University\\
\hspace{-.1cm}$^5$ Drexel University\\
}

\def\d{\mathrm{d}}
\def\be{\begin{equation}}
\def\ee{\end{equation}}

\def\N{\mathrm{N}}

\begin{document}

\date{Draft, \today}

\maketitle

\begin{abstract} 
  We present a simultaneous detection of gravitational magnification
  and dust reddening effects due to galactic halos and large-scale
  structure.  The measurement is based on correlating the brightness
  of $\sim$85,000 quasars at $z>1$ with the position of 20 million
  galaxies at $z\sim 0.3$ derived from the Sloan Digital Sky Survey
  and is used to constrain the galaxy-mass and galaxy-dust correlation
  functions up to cosmological scales.

  The presence of dust is detected from 20 kpc to several Mpc, and we
  find its projected density to follow: $\Sigma_{\rm dust}\sim
  \theta^{-0.8}$, a distribution similar to mass. The amount of dust in galactic halos is found to
  be comparable to that in disks. On large scales its wavelength
  dependence is described by ${\rm R_V}\simeq3.9\pm2.6$, consistent with
  interstellar dust. We estimate the
  resulting opacity of the Universe as a function of redshift and
  find $\langle A_V \rangle \sim 0.03$~mag up to $z=0.5$. This, in
  turn, implies a cosmic dust density of $\Omega_{\rm
    dust}\simeq5\times10^{-6}$, roughly half of which comes from dust
  in halos of $\sim L^\star$ galaxies.

  We present magnification measurements, corrected for dust
  extinction, from which the galaxy-mass correlation function is
  inferred. The mean mass profile around galaxies is found to be
  $\Sigma\sim 30\,(\theta/1\arcmin)^{-0.8}\;h\,{\rm M_\odot\,pc^{-2}}$ up
  to a radius of 10 Mpc, in agreement with gravitational shear
  estimates.
\end{abstract}

\begin{keywords}
dust -- extinction, reddening -- dark matter -- magnification -- 
large-scale structures -- quasars -- galaxies
\end{keywords}

%%%%%%%%%%%%%%%%%%%%%%%%%%%%%%%%%%%%%%%%%%%%%%%%%
\section{INTRODUCTION}\label{sec:introduction}

Light rays from distant sources carry unique information about the
matter and gravitational potential along the line-of-sight. A
well-known example is the signature of intervening gas clouds
imprinted into spectra of background sources via absorption lines.
Mass concentrations located along the path of photons can also induce
gravitational lensing effects.  Background sources can be magnified,
as detected by \cite{2005ApJ...633..589S}, and galaxy shapes can be
distorted as measured through galaxy-galaxy lensing
(cf. \citealt{2002ApJ...577..604H}, \citealt{2004AJ....127.2544S},
\citealt{2005MNRAS.361.1287M}, \citealt{2007ApJS..172..219L,2007ApJ...669...21P}) and
cosmic shear (e.g. \citealt{2008A&A...479....9F}).  Measuring these
effects has become a powerful tool for probing the mass distribution
in the Universe.

In addition, dust extinction effects are expected to occur as
radiation pressure from stars and galactic winds triggered by
supernovae are expected to expel some amount of dust from galactic
disks into the intergalactic medium \citep{1999ApJ...525..583A,2005MNRAS.358..379B}.
Detecting dust reddening by galaxy halos would provide us with useful
information on the life cycles of dust particles as well as
characterize the opacity of the Universe.  In practice, detecting such
an effect is made difficult by the requirement to measure brightness
and/or color changes at a sub-percent level on 100 kpc scales around
galaxies.  A first attempt to find dust in galactic halos was made by
\cite{1994AJ....108.1619Z} who reported a 3-$\sigma$ indication for a
color change of background galaxies found around two nearby spiral
galaxies.  Probing dust reddening induced by galaxy
halos has not been revisited since then, despite the dramatic
improvement in data quality and sample size.

In this work we investigate simultaneously gravitational lensing and
dust reddening of background quasars by foreground
galaxies and associated large scale structure. 
Our observational results primarily make use of the angular
cross-correlation between the brightness of quasars and the projected
density of galaxies.  We first recover and improve upon previous
measurements of the magnification of quasar images by gravitational
lensing due to galaxies located nearby the light paths using a sample
of 85,000 quasars behind 20 million galaxies derived from the Sloan
Digital Sky Survey (\citealt{york2000}; SDSS) Data Release 3
(\citealt{2005AJ....129.1755A}).  Second, this large sample -- together with high
accuracy photometry in five optical pass bands  -- allows us to detect
the presence of dust in the intervening space and explore its
distribution and properties.  This allows us to study the properties
of intergalactic dust and provides a way of inferring the abundance of 
dust in the Universe.\\

We introduce the formalism of brightness-density correlations in
\S2. Data processing and measurements are presented in \S3 \& \S4. The
astrophysical results are given in \S5 and we summarize in \S6.
When needed we use $\Omega_m=0.3$, $\Omega_\Lambda=0.7$
and $H_0=100\;h\;{\rm km\,s^{-1}Mpc^{-1}}$.

% ==============================================================
% ==============================================================
\section{Formalism of Brightness-Density correlations}
\label{sec:mag_shifts}
% ==============================================================
% ==============================================================

Let us begin by considering a background source angularly close to a
foreground galaxy. This galaxy acts as a gravitational lens,
magnifying the source flux and giving rise to extinction and reddening
due to the presence of dust surrounding it.  The observed flux is then
modified by the combination of these effects according to
\be
{\rm f_{obs}} = 
{\rm f_{0}}\,\mu\,{\rm e}^{-\tau_\lambda}\,,
\ee
where $\mu$ is the gravitational magnification and $\tau_{\lambda}$ is the
optical depth for dust extinction at an observed wavelength $\lambda$.
The corresponding magnitude shift is
\be
\delta m_\lambda=-2.5\log\mu\nonumber+\frac{2.5}{\ln 10}\,\tau_{\lambda}\,.\\
\label{eq:main}
\ee
When $\mu$ departs weakly from unity, we can re-express this relation as 
\begin{equation}
\delta m_\lambda \simeq 1.08 \left(\tau_{\lambda}-\delta\mu\right)\,,
\label{eq:delta_m}
\end{equation}
where $\delta\mu=1-\mu$.  Thus, magnification and extinction compete in
changing the brightness of background sources.  Dust extinction is in general
wavelength dependent while magnification is achromatic, so the two effects 
can, in principle, be separated using multi-color data.
Below we show how correlations between the density of foreground 
galaxies and the magnitude of background sources allow us to constrain
the galaxy-mas and galaxy-dust correlation functions.

% ==============================================================
\subsection{The galaxy-mass correlation probed with magnification}
\label{sec:magnification}
% ==============================================================

The galaxy-mass correlation is a powerful tool to probe
the connection between matter and galaxies, providing direct
constraints on the dark matter distribution and models of galaxy formation.
To introduce it, we first define the galaxy and mass overdensities:
\be
\delta_{g}({\bf x})=\frac{n_{\rm g}({\bf x})}{\langle n_{\rm g}\rangle}-1\,
~~~{\rm and}~~~
\delta_{m}({\bf x})=\frac{\rho({\bf x})}{\langle \rho\rangle}-1\,
\ee
where $n_{\rm g}$ and $\rho$ are the density of galaxies and matter.
The galaxy-mass correlation is then defined by
\begin{equation}
{\rm \xi_{gm}}(r)=\langle \delta_{\rm g}(x)\,\delta_m(x+r) \rangle\;.
\label{eq:gm_3d}
\end{equation}
This cross-correlation can be related to the projected surface density of galaxies: 
\begin{eqnarray}
\langle \delta_{\rm g}(\phi)\;\Sigma(\phi+\theta) \rangle &=&
\langle \Sigma(\theta) \rangle 
\nonumber\\
&=& \bar\rho\,\int  \xi_{\rm gm}(\sqrt{\theta^2+\chi^2})\;\d\chi
\label{eq:gm_2d}
\end{eqnarray}
The first relation indicates that the galaxy-mass correlation
is equal to the mean mass profile around galaxies, 
at a given separation $\theta$. The second relation is simply a projection of 
the 3-dimentional galaxy-mass correlation introduced above in 
Equation~\ref{eq:gm_3d} and where $r^2=\theta^2+\chi^2$.

The mass surface density  $\Sigma$ can be probed with gravitational lensing. 
In the weak lensing regime, it is straightforwardly related to the observable
magnification, according to
\begin{equation}
\delta\mu\simeq \kappa/2 =  \Sigma/\Sigma_{crit}.
\end{equation}
Here $\kappa$ is the lensing convergence and the critical mass surface
density is given by
\begin{equation}
\Sigma_{crit}^{-1}=\frac{4\pi\,G}{c^2}\frac{D_d\,D_{ds}}{D_s}\,,
\end{equation}
where $D_l, D_{s}$ and $D_{ls}$ are respectively the angular diameter distances
to the lens, the source and between the lens and the source.

As indicated in Equation~\ref{eq:delta_m}, magnification will affect the 
brightness of background sources and induce a correlation between the density 
of foreground galaxies and the magnitude of background sources.
In order to understand the impact on observable quantities, 
let us consider a given area of the sky and
let $\mathrm{N_0}(m)$ be the intrinsic magnitude distribution of some
sources. The photons originating from these sources may be deflected
by gravitational lensing and magnification affects their
magnitude distribution such that
\begin{equation}
\N(m) \propto \N_0(m-\delta m_{ind})
\label{eq:convolution}
\end{equation}
where $\delta m_{ind}=-2.5\,\log \mu$ is the induced magnitude shift.
This leads to an observable mean magnitude shift:
\begin{equation}
\delta m_\mathrm{obs}=
\left\langle{m}\right\rangle-
\left\langle{m_0}\right\rangle~.
\label{eq:obs}
\end{equation}
It should be noted that for a magnitude-limited sample of sources, the
mean magnitude shift induced by a population of foreground galaxies,
$\delta m_{ind}$, differs from the observable mean magnitude shift of
the individual sources, $\delta m_{obs}$. The difference between the
two depends on the shape of the source magnitude distribution
$\mathrm{N}(m)$ and the limiting magnitude $m_\ell$ of the sample.  In
the case where the induced magnitude shift $\delta m$ is small
compared to the limiting magnitude of the sample, the difference
between the observed and induced magnitude shift can be linearized in
$\delta m_{ind}$ and we have
\begin{eqnarray}
\delta{m_\mathrm{obs}}&\simeq&\mathrm{C_S}\,\times\,\delta m_{ind}\,,
\end{eqnarray}
where the coefficient ${\rm C_S}$ depends on the shape of the magnitude
distribution and the limiting magnitude. In the considered limit, it is given 
by
\begin{equation}
{\rm C_S}= 1-
\frac{1}{\rm N_0^{tot}} \, \frac{\d {\rm N}}{\d m}(m_\ell)\,
\big[m_\ell-\langle m_0 \rangle \big]
\label{eq:Cs}
\end{equation}
(see the derivation in the appendix).  If the sample of sources is not
magnitude-limited, $\d {\rm N}/{\d m}(m_\ell)=0$ and $\mathrm{C_S}=1$,
i.e. the measured magnitude shift equals the induced one.  For
magnitude-limited samples, in general we have ${\rm C_S}<1$, i.e. the
observable magnitude shift is smaller than the induced one. It is
important to note that for power-law luminosity distributions ${\rm
  C_S}=0$. Therefore, while the brightness of each object changes by
$\delta m$, the mean magnitude of the sample remains unchanged.  This
is due to the inclusion of sources which become brighter than the
limiting magnitude. For SDSS quasars in the $g$-band, down to a
limiting magnitude of $g=21$, we find ${\rm
  C_S}\simeq 0.25$.\\

Let us suppose that a population of background quasars and foreground
galaxies are sufficiently well separated in redshift so that physical
correlations between them can be neglected. Gravitational lensing will
give rise to an apparent correlation between quasar brightness and the
proximity to a galaxy overdensity, which we can write as
\begin{eqnarray}
\langle \delta{m_\mathrm{obs}} \rangle(\theta)
&=&
\langle \delta{m_\mathrm{obs}}(\phi)\,\delta_{\rm g}(\phi+\theta) \rangle
\nonumber\\
&\simeq& {\rm C_S}\, \langle \delta m_{ind}(\phi)\,\delta_{\rm g}(\phi+\theta) \rangle
\label{eq:w0}
\end{eqnarray}
where $\delta m_{obs}=m-\langle m\rangle$ is the quasar magnitude
fluctuation and $\delta_{\rm g}$ is the foreground galaxy overdensity
at a given angular distance $\theta$ from a background source. If
magnification effects are in the weak regime,
i.e. $\Sigma/\Sigma_{crit}\ll1$, the above relation reads
\begin{eqnarray}
\langle \delta{m_\mathrm{obs}} \rangle(\theta)
&\simeq& -1.08\,{\rm C_S}\, 
\langle \delta \mu(\phi)\,\delta_{\rm g}(\phi+\theta) \rangle\nonumber\\
&\simeq& -1.08\,{\rm C_S}\;\frac{ \langle \Sigma \rangle(\theta)}{\Sigma_{crit}}\,.
\label{eq:get_mu}
\end{eqnarray}
On scales below a few arcminutes, a non-linear treatment of the magnification 
as a function of the density contrast is needed. Such calculations are presented 
in \cite{2003A&A...403..817M}, where
the authors show that non-linear effects can increase the amplitude of the
magnification by about 30\% at a scale of one arcminute.

For populations of quasars and galaxies selected over an appreciable
redshift range, the above correlation can be expressed in terms of the
galaxy-dark matter cross-power spectrum using the Limbers'
approximation.  Following the formalism introduced by
\citet{1995A&A...298..661B} and the notation laid out in
\citet{jai03}, we can write
\begin{eqnarray}
\langle \delta{m_\mathrm{obs}} \rangle(\theta)
 &=& -1.08\,\mathrm{C_S}  \times
12 \pi^2 \Omega_M\times\\
&&\int \d\chi\,\int k\,\d k~ {\cal K}(k, \theta, \chi)\,
{\rm P_{gm}}(k,\chi)\nonumber
\label{eq:theory_wgq}
\end{eqnarray}
where $\Omega_M$ is the cosmological matter density relative to
critical, $\chi$ is the comoving distance, ${\cal K}$ is the lensing
kernel, and $P_{gm}(k,\chi)$ is the galaxy-dark matter cross-power
spectrum.  The quantity $\langle \delta{m_\mathrm{obs}} \rangle_{\rm
  g}(\theta)$ therefore probes the magnification due to individual
galaxy halos on small scales and on large
scales it constrains the large scale distribution of matter in the Universe.\\

% ==============================================================
\subsection{Dust extinction and reddening}
\label{sec:extinction}
% ==============================================================

Statistical properties of the distribution of dust around galaxies
can be constrained by the galaxy-dust correlation function:
\begin{equation}
\xi_{gd}(r)= \langle \delta_{\rm g}(x)\,\delta_d(x+r) \rangle\;.
\label{eq:gd_3d}
\end{equation}
where $\delta_d(\textbf{x})$ is the dust density fluctuation.  This
cross-correlation can be related to the projected dust surface density
of galaxies:
\begin{eqnarray}
\langle \delta_{\rm g}(\phi)\;\Sigma_d(\phi+\theta) \rangle &=&
\langle \Sigma_d(\theta) \rangle 
\nonumber\\
&=& \bar\rho\,\int  \xi_{gd}(\sqrt{\theta^2+\chi^2})\;\d\chi\;.
\label{eq:gd_2d}
\end{eqnarray}

At optical wavelengths, dust extinguishes and reddens the light of
background sources. The galaxy-dust correlation can then be probed by
measuring the cross-correlation between the colors of background
sources and the distribution of foreground matter.  

Let us define the magnitude shift of a background source measured at a
wavelength $\lambda_\alpha$ with respect to the mean magnitude of the
sample,
\be
\delta m_\alpha({\bf\phi})=m_\alpha({\bf \phi}) - \langle m_\alpha\rangle\,.
\ee
Expanding the formalism introduced in the previous section to
dust reddening effects, and defining the color excess  
\begin{eqnarray}
E_{\alpha\beta}\equiv E(\lambda_\alpha-\lambda_\beta)
&=& \delta m_\alpha-\delta m_\beta\,, \nonumber \\
&\simeq& 1.08 \; [\tau(\lambda_\alpha)-\tau(\lambda_\beta)]  
\end{eqnarray}
we introduce the quasar color-galaxy density correlation, which
can be used to probe the galaxy-dust correlation:
\begin{equation}
 \langle\, {\rm E}_{\alpha\beta}({\bf \phi})\,
\delta_{\rm g}({\bf \phi+\theta}) \,\rangle
\simeq 1.08 \; \langle\, [\tau(\lambda_\alpha)-\tau(\lambda_\beta)]  \,\rangle(\theta)\;.
\label{eq:Edelta}
\end{equation}
If the foreground galaxy and background source populations are well
separated in redshift, the above correlation provides us with
information on the mean reddening and therefore the mean amount dust
around galaxies. The information at different wavelengths can be used
to probe the shape of the extinction curve.  For scales on the order
of a typical galaxy virial radius and smaller, the signal is expected
to be dominated by single galaxies and provides with the average dust
density profile around galaxies. On larger scales it provides
information as to the large scale distribution of dust in the
Universe.

For a given extinction curve the above quantity can be used to infer
the mean dust extinction profile around galaxies $\langle {\rm
  A_\lambda}\rangle (\theta)$, or similarly the mean optical
depth for dust extinction $\langle {\rm \tau_\lambda}\rangle_{\rm
  g}(\theta)$.

It should be noted that not only extinction but also reddening effects
can cause quasars to drop out of the selection criteria as their
identification is a function of colors. In the latter case, the
measured reddening excess turns out to be lower than the true value
and the relation between observed and induced color change needs to be
quantified.  In the present analysis, reddening effects are
sufficiently small for this to be negligible (see discussion below).\\

The measurement of these galaxy-mass (Equation~\ref{eq:gm_2d}) and galaxy-dust 
(Equation~\ref{eq:Edelta}) correlations will be the focus of the remainder
of this paper.

% ==============================================================
% ==============================================================
\section{ANALYSIS}\label{sec:analysis}
% ==============================================================
% ==============================================================

% ==============================================================
\subsection{The data}\label{sec:data}
% ==============================================================

The data set consists of galaxy and quasar catalogues and is drawn
from the third SDSS data release (DR3;~\citealt{aba03}).  
The survey provides images in five broad optical bands ($u,g,r,i,z$ 
\citealt{1996AJ....111.1748F,2002AJ....123.2121S}).
Before masking, this set covers roughly 5300 square degrees, the majority of
which is located around the North Galactic Cap.  
To reduce systematic
errors in the photometric data, we impose a seeing limit of 1''4 and a
Galactic extinction limit of 0.2 in the $r$ band (\citealt{scr02}).
We also apply a mask blocking a one-arcminute radius around bright
galaxies ($r < 16$) and stars with saturated centers to prevent
changes in the source number densities due to local fluctuations
induced by errors in sky brightness subtraction
(\citealt{2005MNRAS.361.1287M}).  Altogether, the masks reduce our
total area to $\sim 3800$ square degrees.

With these cuts, we can reliably perform star/galaxy separation using
Baye\-sian methods to $i = 21$ (\citealt{scr02}), yielding a galaxy
sample whose density is independent of local variations in seeing,
Galactic extinction, sky brightness or stellar density after
masking. This yields about 24 million galaxies between $17 < i < 21$
at a density of approximately 1.8 galaxies per square arcminute.  The
\cite{2005ApJ...633..589S} paper used a sample that was instead r-band
limited, which accounts for the difference in the galaxy sample sizes.

For galaxy magnitudes, we use composite model magnitudes, constructed from
the {\it expMag}, {\it devMag} and {\it fracDev} parameters in the SDSS
database.  This provides a more robust estimate of galaxy flux
at faint magnitudes than the Petrosian magnitudes used in the SDSS
spectroscopic sample, which can have strong variations due to local seeing
for $r > 18$.  For quasars, we consider the flux within the psf profile,
designated as {\it psfMag}.  All magnitudes are de-reddened
to correct for Galactic extinction before applying the various magnitude cuts. 

%-----------------------------------------------------------------------------
\begin{figure}
\begin{center}
  \includegraphics[width=\hsize]{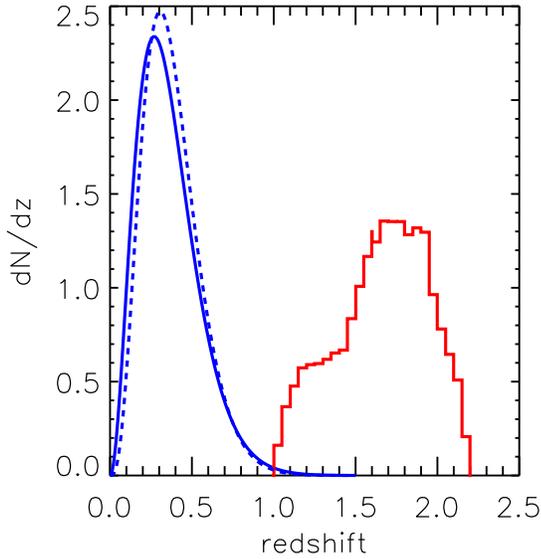}
  \caption{Normalized redshift distributions of the galaxies (solid
    blue line) and photometric QSOs (solid red line). The dashed line
    shows the galaxy redshift distribution weighted by the
    corresponding lensing efficiency.}
\label{fig:redshift_distribution}
\end{center}
\end{figure}
%-----------------------------------------------------------------------------

The quasar data set was generated using the kernel density estimation
(KDE) method described in \citet{ric04} applied to the DR3 data set.
The KDE method is an extension of the traditional color selection
technique for identifying quasars.  Two training sets, one for stars
and one for quasars are prepared and the colors for each object are
compared to those of the two training sets using a 4D Euclidean
distance.  The objects are then classified as either quasar or star
according to a larger probability of membership.  This technique
allows a clean separation of relatively low redshift ($z \leq 2.5$)
quasars from the stellar locus, producing a catalog of 225,000 quasars
down to a limiting magnitude of $g=21$ with efficiency and completeness 
greater than the SDSS spectroscopic target selction algorithm 
(\citealt{ric02, bla03}).  After masking, the total quasar population is 
reduced to 195,000.
\footnote{Upcoming analyses will make use of the larger and deeper
SDSS photometric quasar catalog \citep{2007AAS...21114202R}, 
increasing the number of objects by an appreciable amount.}

In addition to finding quasars, we applied photometric redshift
techniques (\citealt{wei04}) to exclude low redshift quasars which
might be physically associated with our foreground sample.  Quasar
photometric redshifts are driven by the broad spectral emission
features in their spectra, resulting in photometric redshift
likelihoods that can have multiple peaks as well as strong asymmetries
around the most likely redshift.  Rather than estimating a Gaussian
redshift error, we use an upper and lower redshift bounds for a
specified likelihood. To minimize the overlap with the galaxies in the
redshift space, we require the upper and lower bounds to be within the
range $1 < z < 2.2$.  The corresponding distribution of quasar
redshift probabilities is shown in
Figure~\ref{fig:redshift_distribution} with the red line. For
simplicity we treat the redshift p.d.f. of each quasar as a flat
distribution between the upper and lower redshift bounds described
above and weighted by the likelihood that the redshift was within
those bounds. This final selection criterion reduces the number of
quasars to 85704.

%-----------------------------------------------------------------------------
\begin{figure*}
\begin{center}
\includegraphics[width=.48\hsize]{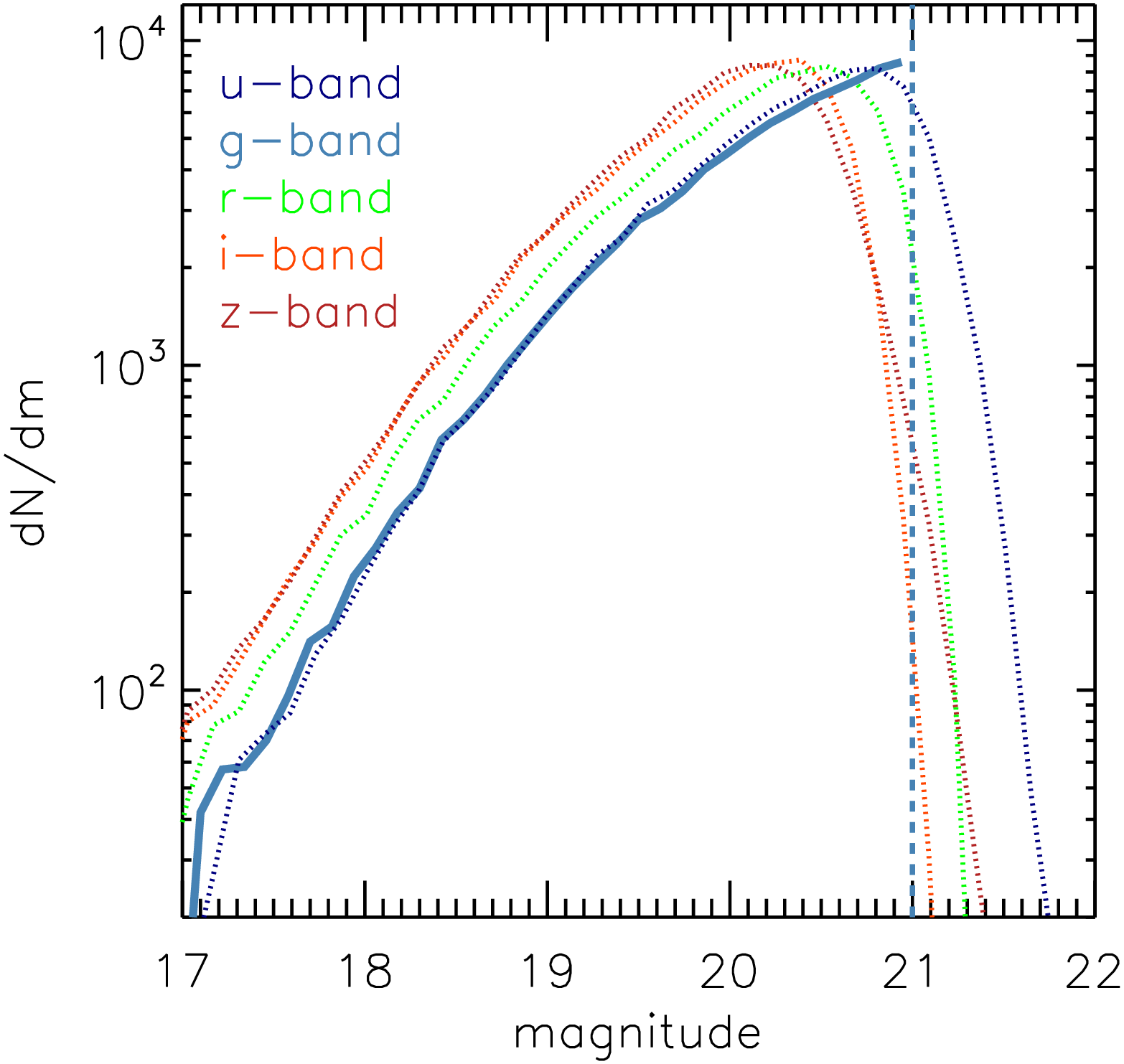}
\includegraphics[width=.48\hsize]{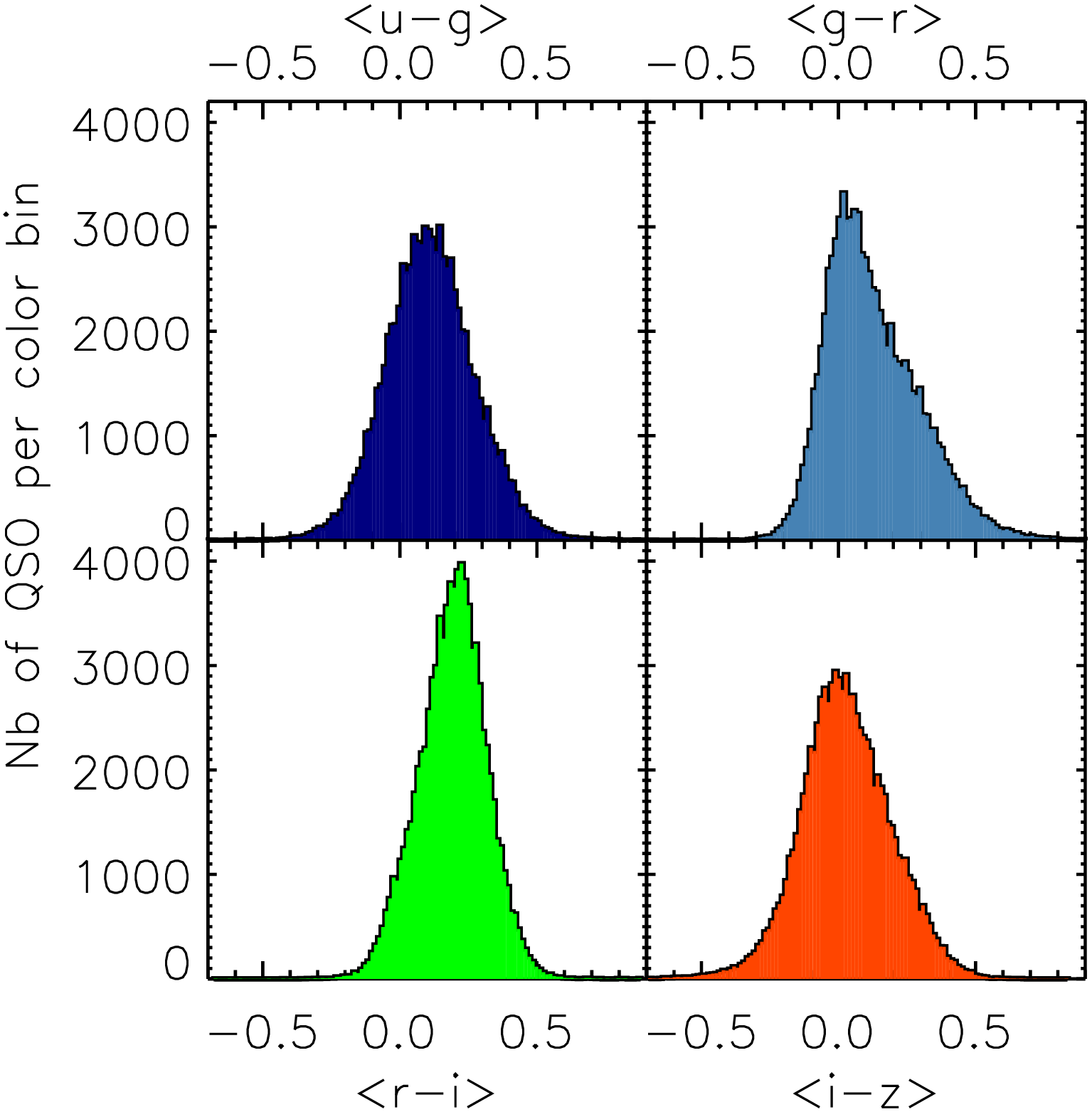}
\caption{\emph{Left: } number counts of our selected sample of photometric quasars 
with $z>1$ as a function of magnitude. 
  The solid line shows the counts in the $g$-band, where
  the quasar sample is magnitude limited.
\emph{Right: } color distributions of the quasars, showing that 
the limiting quasar colors are significantly
greater than the modes of the color distributions. 
Note that for better display 
purposes, we show only objects with colors lower than 0.8}
\label{fig:magnitude_g}
\end{center}
\end{figure*}
%-----------------------------------------------------------------------------

We estimate the overall shape of the galaxy redshift distribution
based on the CNOC2 luminosity functions (\citealt{lin99}) following
the treatment described in \citet{dod02}.  This distribution is well
fit by the expression \be \left( \frac{dN}{dz} \right)_{\rm g}\simeq
z^{2}\,{\rm e}^{-(z /0.187)^{1.26}}\,,
\label{eq:redshift_gal}
\ee
plotted with the solid blue curve. The mean redshift for the galaxy sample
is found to be 
\begin{equation}
\langle z \rangle \simeq 0.36~. 
\label{eq:mean_z}
\end{equation}
For lensing purposes it is useful to compute the galaxy redshift distribution
weighted by the expected lensing efficiency. We have computed this quantity
using the quasar redshift distribution and we show it using the blue dashed
line in the figure. The effective redshift weighted by the lensing efficiency
is given by
\begin{equation}
\langle z_{lens} \rangle
= \frac{\int \d z_g\, \d z_Q \,z_g\;\frac{\d N}{\d z_g}\;\frac{\d N}{\d z_Q}\,\Sigma_{\rm crit}^{-1}(z_g,z_Q)}
{ \int \d z_g\, \d z_Q \;\frac{\d N}{\d z_g}\,\Sigma_{\rm crit}^{-1}(z_g,z_Q)} \;.
\end{equation}
It is found to be $z\simeq 0.38$.

Figure \ref{fig:magnitude_g} shows the quasar magnitude and color 
distributions. As can be seen, the 
quasar sample is magnitude limited in $g$ but not color limited. While quasars 
are selected both in magnitude and color space, the modes of the color 
distributions are far from the limiting colors. As a result, a measured color 
change can be directly used as an estimate of the intrinsic color change of 
the population. Such a property is not satisfied for magnitude changes 
(see section \ref{sec:extinction} for more details).

%-----------------------------------------------------------------------------
\begin{figure*}
\begin{center}
  \includegraphics[height=14cm,width=.99\hsize]{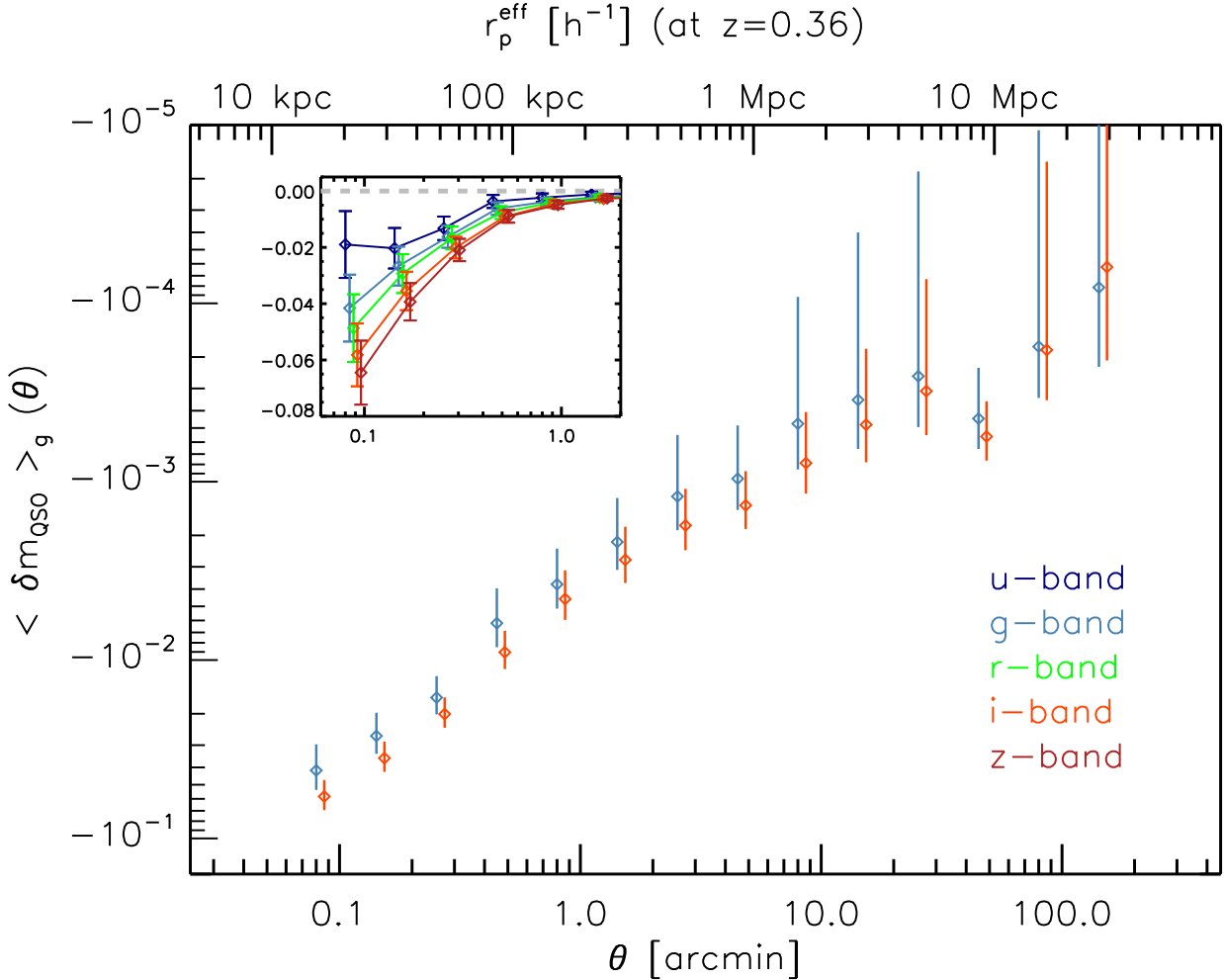}
\vspace{-0.5cm}
\caption{The anti-correlation between quasar magnitude and foreground
  galaxy overdensity due to magnification, as a function of scale. 
  The light blue and orange data points show the measurements in the
  $g$ and $i$ bands and their systematic offset suggests the existence
  of dust extinction. We illustrate its effect in the inset where we show,
for the five SDSS passbands, the wavelength dependence of the signal on small scales.
For reference the physical scale at the mean redshift of the galaxy sample is shown
in the top axis.}
\label{fig:mag_shift_log}
\end{center}
\end{figure*}
%-----------------------------------------------------------------------------

% ===========================================================
\subsection{Measurement}\label{sec:measurement}
% ===========================================================

We measure the density of galaxies (taking into account missing area
due cuts on the local seeing variations, bright stars, etc.) and
compute its correlation, $w_\alpha$, with the magnitude of background
quasars in the band $\alpha$, as a function of angular separation:
\begin{equation}
w_\alpha(\theta)=
\langle \delta{m_\alpha}(\phi)\,\delta_{\rm g}(\phi+\theta) \rangle\;.
\label{eq:w_definition}
\end{equation}

On small scales ($<0.01^\circ$), the signal contributions from
magnification and dust extinction are expected to be dominated by
Poisson noise.  For larger scales, measurements from different angular
bins become significantly correlated, as one would expect since
different quasars will be sampling the same local population of
galaxies.  Since we are looking for very small variations in the
quasar magnitude and galaxy density, the photometric calibration
across the survey needs to be highly homogeneous.  To avoid possible
drift in the SDSS calibration over time, we measure the QSO-brightness
galaxy-overdensity correlation separately on each stripe of the
survey, and then take the average over all stripe-based estimators
\footnote{Adjacent stripes of SDSS data may
  be taken on nights with significant observational lag.  As such, the
  photometric zero points across stripes are not always calibrated to the 
  precision required for this measurement.  In principle, this implies that we 
  should only use data from single scans over one night.  However, in 
  practice, we find that stripe-by-stripe treatment yields a sufficent 
  photometric zero point accuracy to avoid systematic effects on the angular
  scale of the SDSS stripe.
  In the future, using the photometric \"uber-calibration 
  \citep{2008ApJ...674.1217P} might solve some of the
  above issues. }:
\be
w_{j}(\theta) = \frac{1}{ N_{\rm stripe}}
\sum_{i=1}^{N_{\rm stripe}} w_\alpha^{(i)}(\theta)\,,
\ee
where $w_\alpha^{(i)}$ is the QSO-magnitude galaxy-overdensity
correlation measured for stripe $i$ using the filter $\alpha$.  The
associated cost is to reduce the sensitivity of the estimator to power
arising from one direction, i.e., the scanning direction, which is
equivalent to reducing the size of the sample when measuring the
large-scale power.  Our strongest signal is on angular scales smaller
than the width of a single stripe ($2.5^\circ$) so the
extent of this effect should be minimal.\\

In order to measure reddening induced correlations, we compute the
difference between two stripe-based estimators at different
wavelengths,
\be
w_{\alpha\beta}(\theta) = w_\alpha(\theta) - w_\beta(\theta)\,.
\ee
To estimate the error on the above estimators we use $100$ bootstrap
samples of the quasar catalog with which we re-measure the above quantities and estimate
their dispersion.

% ==============================================================
% ==============================================================
\section{RESULTS}\label{sec:results}
% ==============================================================
% ==============================================================

%-----------------------------------------------------------------------------
\begin{figure*}
\begin{center}
  \includegraphics[height=7cm,width=\hsize]{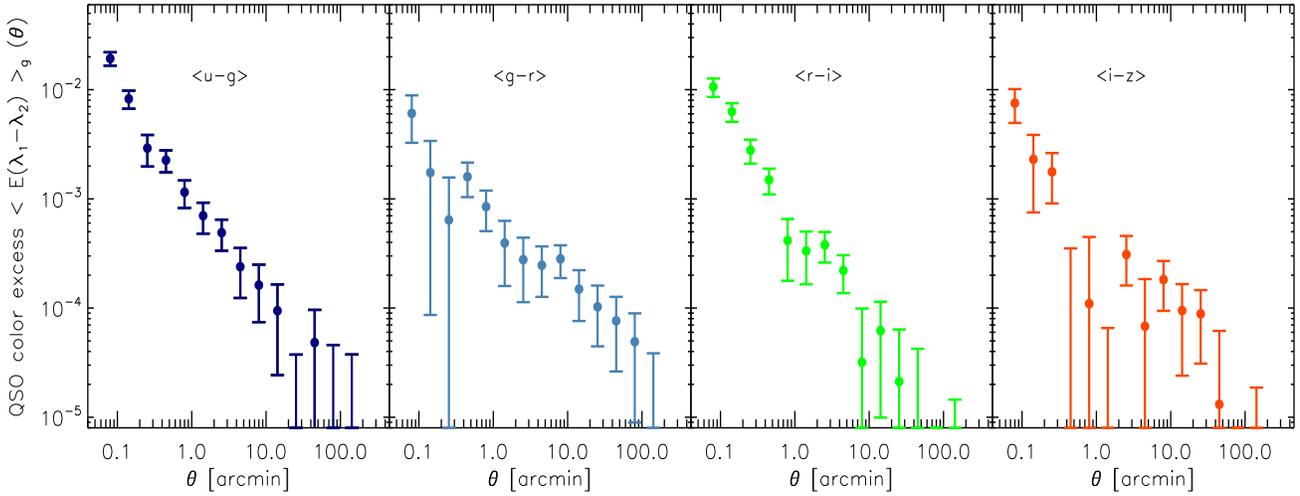}
  \caption{Correlation between QSO reddening and galaxy overdensity as
    a function of angular scale. Note that the four independent
    colors are taken from adjacent passbands and do not
    maximize the signal-to-noise ratio (see Figure~\ref{fig:Av} for
    such a quantity).}
\label{fig:colors_line}
\end{center}
\end{figure*}
%-----------------------------------------------------------------------------

Following the above procedure we measure the correlation between the
observed QSO brightness excess and galaxy density $w_\alpha(\theta)$
(Equation~\ref{eq:w_definition}), where $\alpha$ denotes one of the
five SDSS pass bands, with the errors estimated from bootstrap
resampling.  The results shown in Figure~\ref{fig:mag_shift_log} are
examples for the $g$ and $i$ bands. We observe that quasar magnitude
shifts and galaxy overdensities are anticorrelated.  In other words,
quasars appear to be brighter when closer to galaxies, which implies a
dominance of magnification over extinction effects for the main sample
of SDSS galaxies with $i<21$. At the same time, the systematically
larger amplitude in the redder band indicates the presence of a
wavelength-dependent effect which could be attributed to dust
extinction.

Before interpreting these signals, we note that both magnification and
reddening effects are observed over a wide range of scales, from about
5\arcsec to a few degrees.  We have therefore expanded the angular
range used in the detection of magnification by
\cite{2005ApJ...633..589S} to both smaller and larger separations.
Using the mean galaxy redshift computed in Equation~\ref{eq:mean_z} an
angular scale of one arcminute corresponds to roughly 150 $h^{-1}$
kpc. The above angular range thus corresponds to physical scales from
$\sim20$ $h^{-1}$kpc to $\sim30$ $h^{-1}$Mpc, as indicated in the top abscissa of the
figure.

In the inset of Figure~\ref{fig:mag_shift_log}, we show the wavelength
dependence of the signal on small-scale with measurements of
$w_{\alpha}(\theta)$ with the $u, g, r, i$ and $z$ filters.  We can
observe a continuous trend as a function of wavelength: the signal is
systematically stronger in redder bands, consistent with reddening by
dust.  This also shows that probing magnification requires a
correction for the effects of dust extinction.

%==================================================
%==================================================
\subsubsection*{Systematics Tests}
\label{sec:systematics}
%==================================================

Based on the same quasar sample and a slightly brighter galaxy sample
(selected with $r<21$) \cite{2005ApJ...633..589S} used a density-based
estimator and showed that the dominant magnification signal of the
quasar-galaxy correlation follows the expected dependence as a
function of quasar magnitude. No such behavior could be detected when
using stars instead of quasars as a control sample.

The amplitude of dust reddening depends only on the properties and
amount of dust around the foreground galaxies selected for the
cross-correlation.  We can recover a similar wavelength-dependent
extinction as in Figure~\ref{fig:mag_shift_log} or the mean reddening
signal presented in Figure~\ref{fig:colors_line} by using subsamples
of quasars in different magnitude ranges.  As a further sanity check,
we also replaced our quasars with stars selected to have the same
magnitude and spatial distribution on the sky.  Measuring the
cross-correlation between star brightness and galaxy overdensity with
this sample, we find no appreciable color excess for galaxies with
$i<20.5$ and stars selected in various magnitude ranges.\footnote{ We
  have detected, however, some excess reddening for stars when fainter
  galaxies ($20.5<i<21$) are used. This effect is likely to be
  attributed to a small contamination of faint galaxies in the stellar
  sample at faint magnitudes where star-galaxy classification is
  incomplete.  This is expected to occur at $i\sim21$. The presence of
  galaxies in the star sample may give rise to a similar correlation
  between source reddening and galaxy overdensity, since galaxy
  clustering gives rise to an excess of redder galaxies in overdense
  regions, which produces a similar signal that mimics the reddening
  by dust.  We have measured the reddening-clustering correlation of
  SDSS galaxies with $i>20.5$ and found that a contamination of
  galaxies at about 10\% could explain the reddening signal seen
  around stars.}
Finally, we have investigated the dependence of the signal as a
function of Galactic reddening: by splitting the dataset into two
regimes of Galactic extinction given by the \cite{1998ApJ...500..525S}
map, we did not detect any significant change in our signal.

%==================================================
%==================================================
\subsection{Reddening}\label{sec:reddening}
%==================================================

%==================================================
\subsubsection{Scale dependence}\label{sec:scale}
%==================================================

We isolate the reddening effects by measuring the correlation between
QSO color and foreground galaxy overdensity,
$w_{\alpha\beta}(\theta)$, where $\alpha$ and $\beta$ indicate two
different pass bands. We estimate the errors by computing the color
covariance matrix from bootstrap resampling.  Note that the errors on
colors are smaller than the errors on brightness changes. As
Figure~\ref{fig:colors_line} shows, quasar colors and galaxy
overdensities are positively correlated, i.e.  quasars appear to be
redder when closer to high concentrations of foreground galaxies.  The
reddening effects are detected for all color combinations, from
$\theta\simeq 0.1''$ to about $2^\circ$, corresponding to physical
scales ranging from 50 $h^{-1}$kpc to about 10 $h^{-1}$Mpc.  The
measurement probes galactic radii well beyond the size of galactic
disks.
The amplitude of the effect is stronger in bluer bands.  We find that
a background source whose light passes at around 20 $h^{-1}$kpc
from a foreground galaxy (selected with $i<21$) will be, on average, redder 
by ${\rm E}(g-i)\simeq {\rm E}(B-V) \simeq 0.01$ mag.

%==================================================
\subsubsection{Wavelength dependence}\label{sec:wavelength}
%==================================================

%-----------------------------------------------------------------------------
\begin{figure}
\begin{center}
  \includegraphics[height=9cm,width=\hsize]{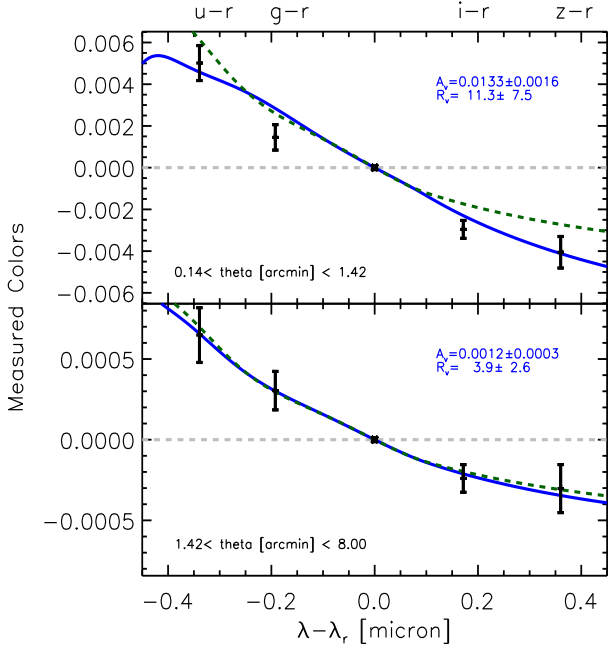}
  \caption{Mean color excess with respect to the r-band measured in
    two angular bins.  The blue curves show the best fit extinction
    curve parametrized by \citet{1994ApJ...422..158O}. 
     The dashed green curves are fits with $R_V=3.1$. Note that $R_V$ is poorly constrained on small scales.}
  \label{fig:color_one_arcminute}
\end{center}
\end{figure}
%-----------------------------------------------------------------------------

The five SDSS filters allow us to constrain the shape of ${\rm A(\lambda)}$, the
extinction curve of the dust associated with the galaxies, through four
independent colors.  We measure the $w_{\alpha\beta}$ correlations for two
angular bins: $0.14<\theta<0.80$ arcmin and $0.8<\theta<8.0$ arcmin,
which correspond to effective projected radii of about $20<r_p<100$
$h^{-1}$kpc and $100\,h^{-1}{\rm kpc}<r_p<1\,h^{-1}$Mpc. In
Figure~\ref{fig:color_one_arcminute} we show the corresponding color
excesses with respect to the $r$ band.
We compare these reddening measurements to the standard extinction
curve by fitting these data points with the functional form of the
extinction curve provided by \cite{1994ApJ...422..158O}.  Such
extinction curves are usually characterized by the parameter
$R_V=A_V/E(B-V)$, which characterizes the slope of the extinction
curve. The coefficient $A_V$ quantifies the amount of dust through its extinction in the $V$ band. The best fit for $A_V$ and $R_V$ is shown with the blue curve. 
On small scales our measured reddening corresponds to
$A_V=(1.3\pm0.1)\times10^{-2}$ mag and $R_V=11.3\pm7.5$, i.e. the slope
of the extinction curve is not well constrained. On large
scales however we obtain a better accuracy:
\begin{equation}
R_V=3.9\pm2.6.  
\end{equation}
and $A_V=(1.2\pm0.3)\times10^{-3}$ mag.
The green dashed curve shows the best fit for $A_V$
when $R_V$ is assumed to be 3.1, the standard value for dust in the disk
of our Galaxy. Within the errors our results are consistent with standard interstellar dust.

%==================================================
\subsection{Dust extinction}
\label{sec:profile}
%==================================================

Given our measurements of the reddening due to foreground galaxies, we can 
now make estimates of the average dust extinction profile of the galaxies in 
our sample.  In addition to constraining the amount and spatial distribution 
of dust on large scales around galaxies, this quantity is of interest to 
quantify the intrinsic brightness of background sources, which is the main 
goal of Type Ia supernovae measurements aimed at constraining dark energy.
%-----------------------------------------------------------------------------
\begin{figure*}
\begin{center}
  \includegraphics[width=0.6\hsize]{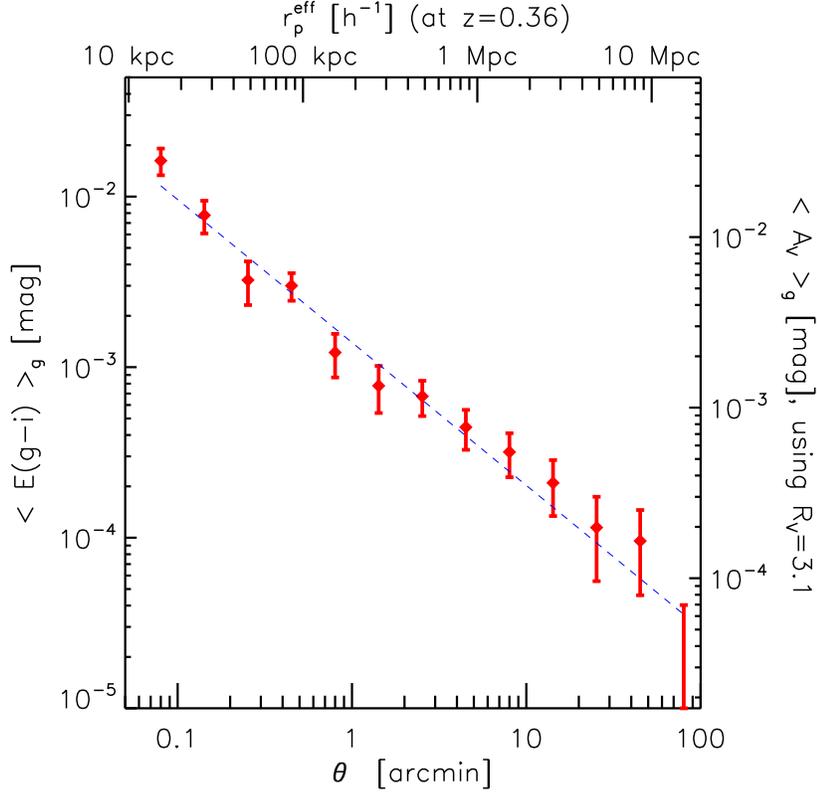}
  \caption{Mean observer-frame $\rm{E}(\emph{g-i})$ reddening profile around galaxies with
    $i<21$, as a function of impact parameter in angular scales (lower
    axis) and effective projected radius (upper axis). Note that we have ${\rm
      E}(g-i)\simeq{\rm E}(B-V)$. The right axis shows the
    corresponding observer-frame extinction in the $V$ band using the standard
    interstellar value $R_V=3.1$.
  }
\label{fig:Av}
\end{center}
\end{figure*}
%-----------------------------------------------------------------------------

As indicated in \S\ref{sec:extinction}, the correlation between quasar
color and galaxy overdensity is an estimator of the mean reddening
induced by galaxies.  In order to maximize the signal-to-noise ratio
of the reddening detection, we measure the quantity
$w_{\alpha\beta}(\theta)$ for which $\alpha$ and $\beta$ are taken to
be the $g$ and $i$ pass bands.  We avoid using the $u$ and $z$ bands
for which the photometric errors are substantially larger.
In addition, the latter suffers from photometric contamination due to
sky emission lines. We show the mean $E(g-i)$ color excess of quasars
as a function of scale in Figure~\ref{fig:Av}. A best fit power-law gives
\begin{equation}
\langle E(g-i)_{\rm }\rangle (\theta) =
(1.4\pm 0.1)\times10^{-3}\,\left(\frac{\theta}{1\arcmin}\right)^{-0.84\pm 0.05}\,.
\label{eq:Egi}
\end{equation}
This quantity is proportional to the dust surface density and
therefore provides us with direct constraints on the spherically
average distribution of dust around galaxies.  The angular dependence
is similar to average mass profiles around galaxies constrained from
galaxy-galaxy lensing
(e.g. \cite{2004AJ....127.2544S,2005MNRAS.362.1451M}) and
magnification measurements by \cite{2005ApJ...633..589S}.
This result has a number of implications regarding the amount and
nature of the dust in galaxy halos. It will be discussed below
in section \ref{sec:implications}.\\

In order to convert reddening into extinction, we choose the value $R_V=3.1$,
corresponding to standard interstellar dust in our Galactic disc and in
agreement with the constraints obtained above.  In Figure~\ref{fig:Av},
we plot the mean extinction profile around the galaxies in our sample.
It can be written as
\begin{equation}
\langle {\rm A_V}\rangle (\theta) =
(2.4\pm0.2)\times10^{-3}\,\left(\frac{\theta}{1\arcmin}\right)^{-0.84\pm0.05}
\label{eq:Av}
\end{equation}
or
\begin{equation}
  \langle {\rm A_V}\rangle (r_p) =  (4.14\pm0.19)\times10^{-3}\,\left(\frac{r_p}{100\,h^{-1}\,{\rm kpc}}\right)^{-0.84\pm0.05}
\label{eq:Av2}
\end{equation}
where $A_V$ is the \emph{observed} $V$-band extinction. 
For an extinction curve characterized by ${\rm R_V}=3.1$, we have
${\rm A}_\lambda\propto \lambda^{-1.2}$ in the visible range.
We point out that the amount of reddening profile shown in Figure \ref{fig:Av}
is significantly lower than contributions expected from dwarf galaxies.
For example, the average dust reddening induced by the LMC, located at about
50 kpc from our Galaxy, is
${\rm E(B-V)}\simeq 0.075\;{\rm mag}$
\citep{1998ApJ...500..525S}. Such a value is about an order of magnitude larger than
eq. \ref{eq:Egi} and shows that satellite galaxies are expected to dominate the amount
of dust reddening for individual lines-of-sight intercepting galaxy halos.

%==================================================
\subsection{Magnification}
%==================================================

%-----------------------------------------------------------------------------
\begin{figure*}
\begin{center}
  \includegraphics[width=.8\hsize]{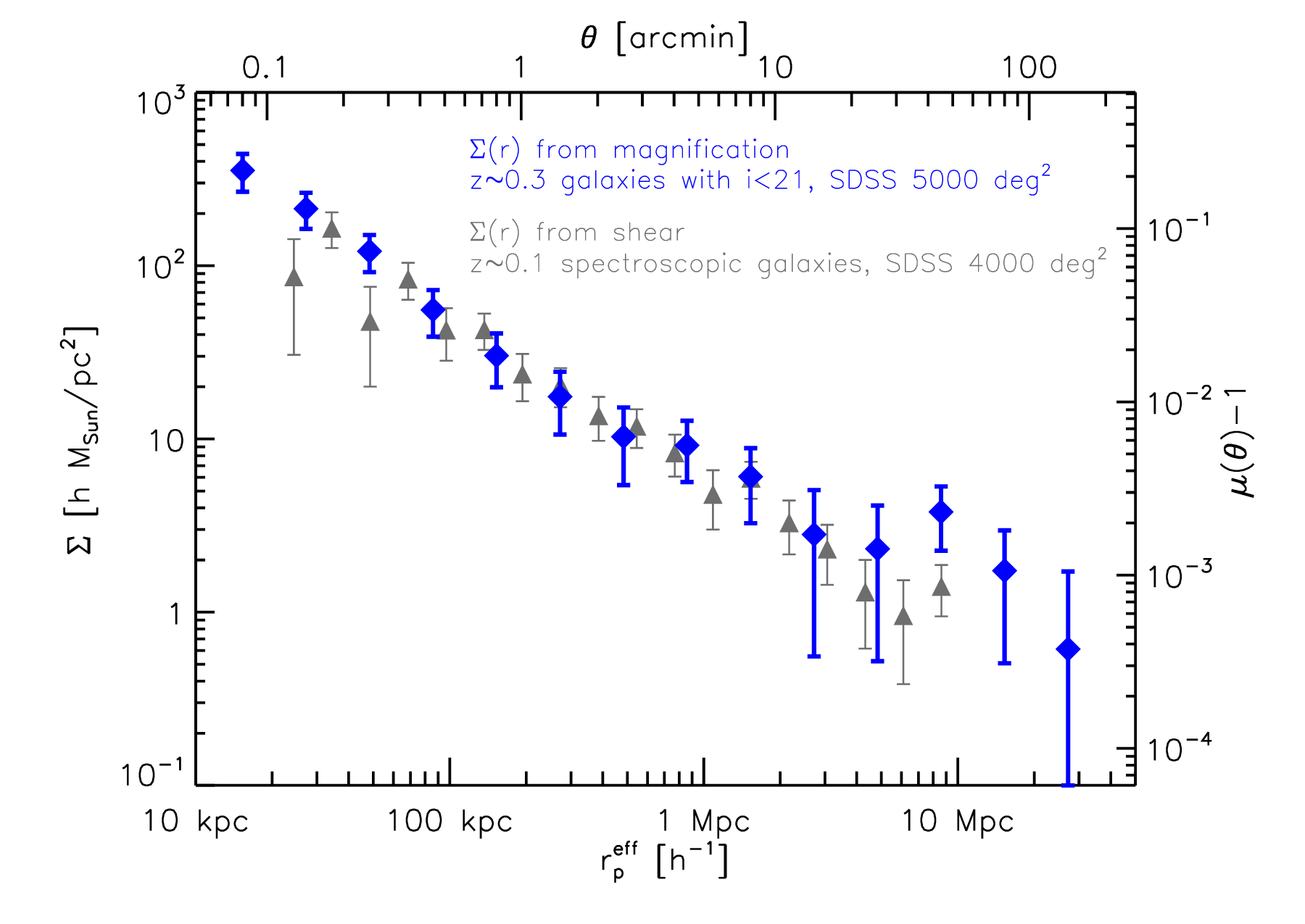}
  \caption{ The mean surface density of
  galaxies (with $i<21$) measured through the magnification of background quasars and corrected for dust extinction (blue points). In comparison we show the mean surface density of a sample of $\sim L^\star$ galaxies at $z\sim0.1$ obtained from the gravitational shear of background galaxies from \citet{2004AJ....127.2544S}. Non-linear magnification effects have
  not been included and result in an overestimation of the mass on the smallest scales.   }
\label{fig:mu_gamma}
\end{center}
\end{figure*}
%-----------------------------------------------------------------------------

The correlation between quasar magnitude shift and galaxy overdensity
allows us to constrain a combination of magnification and dust
extinction.  Having estimated the wavelength-dependent extinction from
reddening measurements in the previous section, we can now use this
information and isolate the magnification effects.
Our goal here is to estimate the signal well enough to allow for a
comparison with similar measurements from galaxy-galaxy lensing
available in the literature. 

For simplicity we start by describing the angular correlation between quasar 
brightness and galaxy overdensity by a simple power-law with index $-0.8$.
As above, we use a value of $R_V=3.1$ (or ${\rm A}_\lambda\propto \lambda^{-1.2}$
in the visible range) to convert reddening 
into extinction. The mean quasar brightness change as a function of angular 
separation from galaxies then reads
\begin{equation}
\langle \delta m_{\rm obs} \rangle(\theta)
\simeq \left [ \delta m_\mu + \delta m_{\tau_V} 
\,\left(\frac{\lambda}{\lambda_V}\right)^{-1.2} \right ]
\,\left(\frac{\theta}{1\arcmin}\right)^{-0.8}\;,
\label{eq:fit_delta_m_obs}
\end{equation}
where $\delta m_\mu$ is the achromatic magnitude change due to magnification
and $\delta m_{\tau_V}$ is the change due to dust extinction in the V-band.
As shown above, we have $\delta m_{\tau_V}\simeq +2.4 \times 10^{-3}$ (Eq. \ref{eq:Av}),
which gives $\delta m_\mu\simeq -6.7\times 10^{-3}$. 

At sufficiently large wavelength, dust extinction
becomes negligible and the brightness change is only due to
magnification. In this limit, we find
\begin{equation}
\langle \delta \mu\rangle (\theta) \simeq
2.5\times10^{-2}\,\left(\frac{\theta}{1\arcmin}\right)^{-0.8}
\label{eq:delta_mu}
\end{equation}
where we have used the conversion between observed and induced
magnitude shift (Equation~\ref{eq:get_mu}) and ${\rm C_S}\simeq 0.25$ for
SDSS quasars with $g<21$ (Equation~\ref{eq:Cs}).  This implies that the
mean magnification excess found at one arcminute (or $\sim$100 $h^{-1}\,$kpc) is
about 2\%. This result is in agreement with that of
\cite{2005ApJ...633..589S} where the authors inferred the
magnification from the observed correlation between quasar and galaxy
densities on the sky, i.e. a different estimator.  Our analysis also
allows us to measure magnification effects on smaller scales.
On the smallest scale we can probe, i.e. 5\arcsec (or $\sim$15 $h^{-1}\,$kpc), 
we find $\delta\mu\simeq 15\%$.\\

In the weak lensing regime, the magnification is related to the mean
mass surface density by
\begin{equation}
\Sigma = {\Sigma_{\rm crit}}\,\frac{1+\delta \mu}{2}\,.
\label{eq:magnification_linear}
\end{equation}
The inferred mean mass surface density of our galaxy sample is shown
with the blue data points in Figure~\ref{fig:mu_gamma}. A best fit power-law
distribution gives
\begin{equation}
{\rm \Sigma}(\theta)
\simeq A\,\left ( \frac{\theta}{1\arcmin} \right)^{-0.8}
\end{equation}
with $A=30.6\pm3.4\,h\,M_\odot\,{\rm pc}^{-2}$, or using our
effective projected scale:
\begin{equation}
{\rm \Sigma}(r_p)
\simeq A'\,\left ( \frac{r_p}{1\,h^{-1}{\rm Mpc}} \right)^{-0.8}
\end{equation}
with $A'=8.1\pm0.9\,h\,M_\odot\,{\rm pc}^{-2}$.  These results provide
an estimate of galaxy density profiles from gravitational
magnification corrected for dust extinction effects.  
 While a
linear relation between magnification and density contrast is a good
approximation on large scales, higher-order corrections become
significant on scales smaller than a few arcminutes. As shown by
\cite{2003A&A...403..817M}, using Eq. \ref{eq:magnification_linear}
results in overestimating the projected mass $\Sigma$ by about 15 to
25\% on scales ranging from 1 to 0.1 arcminute. The first few points
shown in Figure \ref{fig:mu_gamma} have not been corrected for this
effect. As shown below, on larger scales the mass estimate from
magnification is in good agreement with shear-based mass estimates.

% ------------------------------------------------------------------------
\subsubsection{Comparison with shear measurements}
% ------------------------------------------------------------------------

So far the galaxy-mass cross-correlation has mostly been accessible through
galaxy-galaxy lensing measurements which estimate the mean 
tangential shear of background galaxies for a given sample of foreground
lenses (e.g. \citealt{2002ApJ...577..604H}, \citealt{2004AJ....127.2544S}, 
\citealt{2005MNRAS.361.1287M}, \citealt{2007ApJS..172..219L}).
The tangential shear ($\gamma_t$) azimuthally averaged
over a thin annulus at projected radius R from a lens galaxy is directly
related to the projected surface mass density of the lens within the aperture, 
\begin{eqnarray}
\gamma_t = \frac{\Delta\Sigma(R)}{\Sigma_{\rm crit}}
\end{eqnarray}
where
\begin{equation}
\Delta\Sigma(R)=\bar \Sigma(<R)-\Sigma(R)\,,
\end{equation}
$\bar\Sigma(< R)$ is the mean surface density within radius $R$, and
$\Sigma(R)$ is the azimuthally averaged surface density at radius
$R$ (\citealt{1991ApJ...380....1M, 1994ApJ...437...56F}).
Therefore, shear measurements constrain $\Delta\Sigma$ whereas magnification
is a direct estimate of $\Sigma$.

In order to compare magnification and shear measurements, it is interesting to
note that, in the case of power-law mass profiles, with
$\Sigma(r_p)\propto r_p^{-\alpha}$, we have
\begin{equation}
\Delta\Sigma(r_p) = \frac{\alpha}{2-\alpha}\,\Sigma(r_p)\,.
\label{eq:rescale_shear}
\end{equation}
The two observables $\Sigma$ and $\Delta\Sigma$ are therefore equivalent for
isothermal profiles. For an angular dependence following $r_p^{-0.8}$, 
we have $\Delta\Sigma=\Sigma\times f$ with $f\simeq0.66$. This value goes
down to $0.54$ for an index of $-0.7$.
Using galaxy-galaxy measurements based on the SDSS,
\cite{2004AJ....127.2544S} found ${\rm \Delta\Sigma}(R) = A'\,\left (
  r_p/1\,h^{-1}{\rm Mpc}\right)^{-\alpha'}$ with
$A'=(3.8\pm0.4)\,h\,M_\odot\,{\rm pc}^{-2}$ and $\alpha'=0.76\pm0.05$
for a sample of spectroscopically identified lenses with $\langle z
\rangle \simeq0.1$ and $\langle L \rangle \simeq L^\star$.
Considering for simplicity a slope of $-0.8$ (consistent with their
constraint), their results translate into ${\rm \Sigma}(R) =
(5.7\pm0.6)\,\left ( r_p/1\,h^{-1}{\rm Mpc} \right)^{-0.8} \,h\,M_\odot\,{\rm
  pc}^{-2}$.  Their scaled-measurements (using
Equation~\ref{eq:rescale_shear}) are shown in
Figure~\ref{fig:mu_gamma}.  Under these assumptions, the magnification
and shear estimators appear to be in very good agreement.  A more
accurate comparison between the two would require accounting for the
differing luminosity distributions between the two lensing samples as
well possible redshift evolution (our lenses comprising most of the
sources for the shear-based estimators).  However, it illustrates that
both methods offer comparable degree of measurement precision and
dynamic range while affected by significantly different systematic
effects.

%==================================================
%==================================================
\section{IMPLICATIONS}
\label{sec:implications}
%==================================================
%==================================================

Our detection of the change in apparent magnitude of distant quasars has 
allowed us to quantify the magnification and reddening of background sources
as their light rays pass in the vicinity of foreground galaxies.
While the amplitude of gravitational lensing effects has been known both
from theory \citep{1995A&A...298..661B,2002A&A...386..784M,jai03}
and observations \citep{2005ApJ...633..589S}, the expected amplitude 
of dust reddening and extinction effects was largely unconstrained.
Our study has shown that, in the visible bands, dust extinction occurs at a
level comparable to that of the observed brightening due to magnification. 
Hence, studies aimed at predicting observable 
magnitude changes of  background sources (quasars, galaxies, 
supernovae, etc.) which include only gravitational lensing effects are 
incomplete. In the visible range, dust 
extinction effects cannot be neglected and must be included alongside with
magnification to properly account for the effects of passing through
large scale structure.\\

Below, in order to simplify the discussion we will assume the dust in
galactic halos to be described by SMC type dust.
This choice is motivated by a number of results: (i) certain
low-ionisation absorbers such as MgII are known to inhabit the halo of
$\sim L^\star$ galaxies. They are found on scales reaching up to about
100 $h^{-1}\,$kpc around galaxies \citep{2007ApJ...658..161Z}.  Recently several
authors studied their extinction properties and dust content
\citep{2005pgqa.conf...86M,2005pgqa.conf..427K,2006MNRAS.367..945Y}.
They found that their average extinction curve is similar to that of
the SMC, i.e. does not show the 0.2 $\micron$ bump seen in the Milky
way extinction curve (for systems with $z>0.9$ where the feature
enters the visible window).  (ii) In addition, it is known that only
a small fraction of high redshift galaxies show an extinction
curve with the $0.2 \micron$ bump. As discussed below, considering a Milky-Way type dust changes our results only by a factor two.\\

In order to characterize the population of galaxies responsible for
most of the intergalactic dust, we first assume that the amount of dust in 
halos is, on average, proportional to the metallicity and luminosity of galaxies. Under this assumption, 
the dominant contribution of the reddening signal  is expected to originate 
from galaxies with an effective luminosity: 
\begin{eqnarray} 
L_{\rm eff} &\equiv& \int_{L_{min}}^{\infty} \,{\phi}(L)\,{\rm Z}(L)
\,L\,\d L ~/~ \int_{L_{min}}^{\infty} \d L\,\phi(L)\,{\rm Z}(L)
\label{eq:L_mean}
\end{eqnarray}
where ${\rm \phi}(L)$ is the galaxy luminosity density given by the
Schechter function
\begin{equation}
\phi(L)\,\d L = \phi^\star\,\left(\frac{L}{L^\star}\right)^\alpha
{\rm e}^{-L/L^\star}\,\d\left(\frac{L}{L^\star}\right)
\end{equation}
with $\alpha=-1.1$ and ${\rm Z}(L)$ is the metallicity-luminosity
relation \citep{2004ApJ...613..898T}:
\begin{equation}
12+\log \left ( \frac{O}{H} \right) = -0.185\, M_B + 5.328
\end{equation}
with $M_B^\star=-19.5$.
By integrating Equation~\ref{eq:L_mean} between $L=10^{-2}\,L^\star$ and
infinity, we find that the
most important contribution of reddening originates from galaxies with a luminosity
\be L_{\rm eff}\simeq0.45\, L^\star\,.
\label{eq:def_L}
\ee
Given the number density of $L^\star$ galaxies \citep{2006ApJ...639..590F} and the above
luminosity function, galaxies with $L=L_{\rm eff}$ are expected to have  a comoving number density of $n\simeq 0.037\,h^{3}\,{\rm Mpc}^{-3}$.\\

Below we discuss a number of implications given by the existence of a 
large-scale distribution of dust around galaxies.  
We note that only certain of the following results will make use of the above numbers.

%==================================================
\subsection{Dust mass distributions}
%==================================================

The properties of dust particles giving rise to the observed optical
extinction and infrared emission in our Galaxy, the LMC and the SMC
have been modeled by several authors
(e.g. \citealt{2000JGR...10510269M} and
\citealt{2001ApJ...548..296W}). 
These models provide estimates of the
absorption cross section per mass of dust
as a function of wavelength $K_{\rm ext}(\lambda)$, which is used to
infer dust mass surface density from reddening profiles.
For the SMC type dust the model of 
\cite{2001ApJ...548..296W} gives
\footnote{We thank Joseph Weingartner for having provided us with this
value.}
\begin{equation}
\rm{K_{\rm ext}}(\lambda_V)\simeq 1.54\times10^{4}\,\rm{cm}^2\,\rm{g}^{-1}\,.
\label{eq:K}
\end{equation}
Note that Milky Way type dust corresponds to a dust mass larger by a
factor two at a fixed extinction.

%==================================================
\subsubsection{Evidence for a diffuse component of dust in galaxy halos}
%==================================================

\begin{figure}
\begin{center}
  \includegraphics[width=1.\hsize]{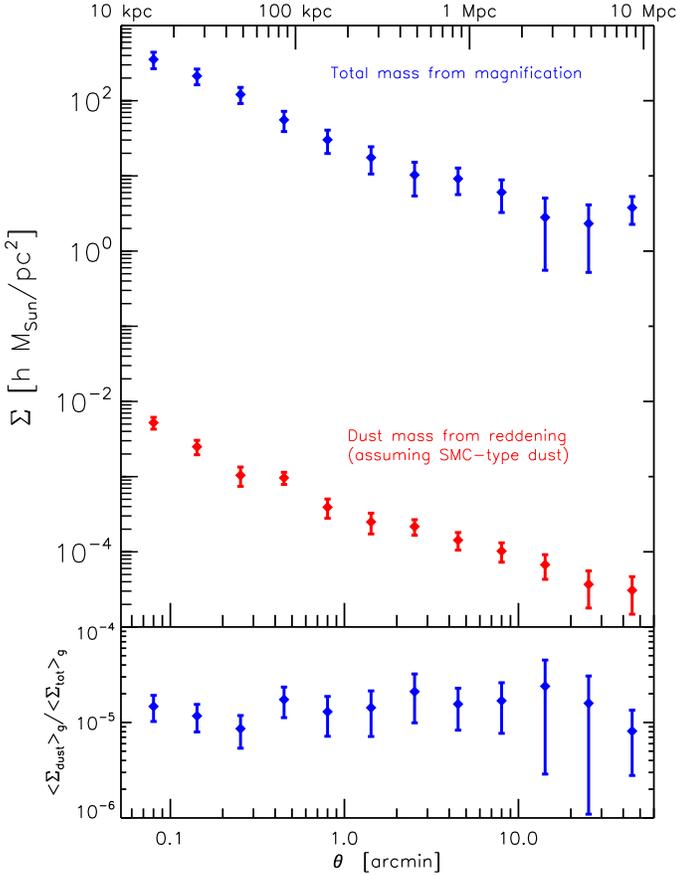}
  \caption{Correlation between magnification and galaxy overdensity as
  a function of scale. On scales smaller than $\sim 500$ $h^{-1}\,$kpc, this
  quantity is a direct estimate of the mean surface density of
  galaxies (with $i<21$).}
\label{fig:mass_profiles}
\end{center}
\end{figure}

The knowledge of $K_{\rm ext}$ allows us to convert the observed reddening into a 
dust mass surface density. The spatial dependence of this quantity is shown in 
Figure~\ref{fig:mass_profiles}. We find a column density of dust of about 
$10^{-3} h\,{\rm M_\odot\,pc^{-2}}$ at an impact parameter of 100 $h^{-1}\,$kpc.  
It is interesting to estimate the total amount of dust in the halo of
the galaxies defined by Equation~\ref{eq:def_L}.  
Considering an isothermal sphere mass distribution, the 
virial radius of $0.5\,L^\star$ galaxies is $r_v\simeq  110\,h^{-1}{\rm kpc}$.
The mass of dust residing in the halo of such galaxies is given
by integrating the ratio  $\mathrm{A_V}(r_p)/K_{\rm ext}(\lambda_V)$
over the area enclosed by $r_v$:
\begin{equation}
  M_{dust}=\frac{2\pi\,\ln 10}{2.5\;{\rm K_{\rm ext}(\lambda_V)}}\,\int_0^{r_{vir}} \mathrm{A_V}(r_p)\,r_p\,\d r_p\;.
\end{equation}
Our measurements provide us with the radial dependence of $A_V$ for
$r_{\rm eff} \gtrsim 20$ $h^{-1}\,$kpc, i.e. scales greater than galactic disks. By 
defining the dust mass in the halo as 
$M^{\rm halo}_{ dust} \equiv M_{dust}(20\,h^{-1}\,{\rm kpc}<r_{\rm eff}<r_v)$
and using eq. \ref{eq:Av} with SMC type dust, we find 
\begin{equation}
M^{\rm halo}_{dust}\simeq 5\times10^7 \,M_\odot\;.
\end{equation}
%-----------------------------------------------------------------------------
%-----------------------------------------------------------------------------
This estimate shows that a substantial amount of dust exists in the halo of 
$\sim L^\star$ galaxies.  
This dust mass is comparable to that commonly found in galactic disks
\citep{2007ApJ...663..866D}.
This immediately shows that the reddening profile shown in Figure \ref{fig:Av}
cannot be explained by a contribution from satellite galaxies whose luminosities
and therefore dust masses is negligible compared to a central $L^\star$ galaxy.
Our results suggest the existence of a diffuse component of dust in galactic halos.

We note that similar arguments can be obtained from reddening considerations
and do not rely on the specific value of $K_{\rm ext}$ (Eq. \ref{eq:K}).
It should also be noted that while the total amount of dust appears to be dominated
by a diffuse component, the largest reddening values are still expected to
originate from lines of sight passing through satellites. The average
reddening value of the LMC, ${\rm E(B-V)}\simeq 0.075\;{\rm mag}$, is about
one order of magnitude larger than the average halo reddening at a radius on
order 50 $h^{-1}\,$kpc (see Figure~\ref{fig:Av}).

%==================================================
\subsubsection{Dust-to-mass ratio}
%==================================================

Having obtained constraints on both the total mass distribution from
magnification and dust mass from reddening (see Figure
\ref{fig:mass_profiles}), we can compare their statistical spatial
distributions by computing the ratio
\begin{equation}
\Gamma(\theta)=\frac
{\langle \Sigma_{\rm dust}(\theta) \rangle}
{\langle \Sigma(\theta) \rangle}\;,
\end{equation}
where $\Sigma_{\rm dust}$ and $\Sigma$ are the dust mass
and total mass surface densities. This ratio is plotted as a function of 
scale in the lower panel of
Figure \ref{fig:mass_profiles} and appears to be only weakly
scale-dependent. We find the dust to mass ratio
\begin{equation}
\Gamma \simeq 1.1\times10^{-5}\;.
\end{equation}
for SMC type dust.

%==================================================
\subsubsection{The cosmic density of dust}
%==================================================

We now attempt to estimate the cosmic density of dust,
$\Omega_{dust}$.  To do so we first compute the density of dust
originating from galaxy halos.  Considering $\sim L^\star$ galaxies
with a mass-to-light ratio ${M_v}/{L_B}=250\,h~M_\odot/L_\odot$
\citep{2004ApJ...616..643F}, a light density ${\cal L}_B\simeq2\times10^8\;h\,{\rm L_\odot\,Mpc}^{-3}$  \citep{2003ApJ...592..819B} and
for a dust-to-mass ratio $\Gamma$ in halos, we can write
\begin{eqnarray}
\Omega_{\rm dust}^{halo} &\simeq& 
\frac{\Gamma\,\times\,({M_v}/{L_B})\,\times\,{\cal L}_B}{{\rho_{\rm crit}}}
\nonumber\\
&\simeq& 2.8\times10^{-6}\;.
\end{eqnarray}
Previous attempts to obtain observational constraints on the cosmic
density of dust have focused on light-weighted estimates, i.e. dust
related to disks.  For example, using the attenuation-inclination
relation for galaxy discs and their associated central bulges,
\cite{2007MNRAS.379.1022D} quantified the mean attenuation of a large
sample of galaxies and estimated $\Omega_{\rm dust}^{\rm
  disk}\simeq2\times10^{-6}$.  Similarly, theoretical estimates have
focused on dust associated with cold gas. For example,
\cite{2004ApJ...616..643F} estimated the cosmic dust density by
computing the mean metallicity of galaxies weighted by the Schechter
luminosity function, used a mass fraction of metals into dust grains
of $Z({\rm dust})/Z=0.2$ multiplied by the density parameter in cool
gas. They find $\Omega_{\rm dust}^{\rm disk}\simeq2.5\times10^{-6}$.
The total cosmic dust density can be estimated by summing up
the halo and disk contributions. We find that
\begin{eqnarray}
\Omega_{\rm dust}
&=&\Omega_{\rm dust}^{halo}
+\Omega_{\rm dust}^{disk}\nonumber\\
 &\simeq& 5.3\times10^{-6}\;.
\end{eqnarray}
This value is in agreement with the (model-dependent) upper limit
obtained by \cite{2004MNRAS.350..729I}: $\Omega_{\rm dust}< 10^{-5}$
at $z\sim0.3$. It is about a factor two larger than the estimate
given by \cite{2004ApJ...616..643F} for galactic disks.

%==================================================
\subsection{The opacity of the Universe}
%==================================================

As our results provide us with an estimate of the mean optical depth
for dust extinction due to a population of galaxies spanning a given
redshift range, we can attempt to integrate $\tau_{\rm g}(z,\lambda)$ over
redshift and estimate the opacity of the Universe.

%-----------------------------------------------------------------------------
\begin{figure}
\begin{center}
  \includegraphics[width=\hsize]{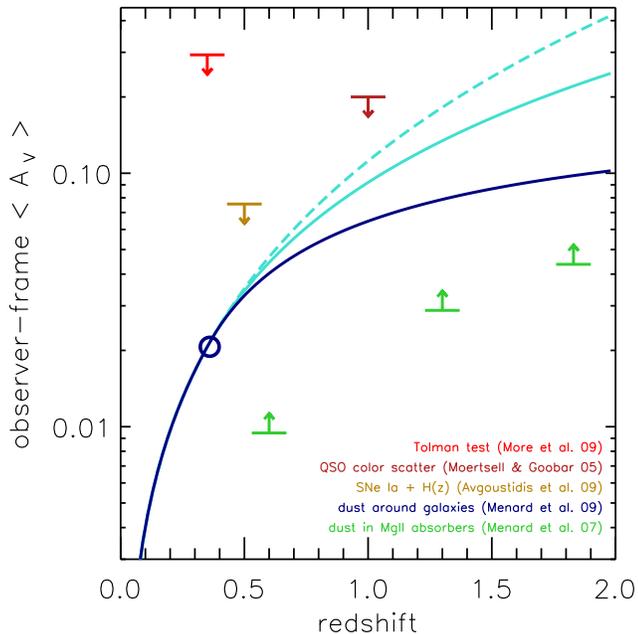}
  \caption{The average observer-frame $A_V$ extinction as a function of
  source redshift. 
  The blue circle shows our (model-dependent) estimate of the opacity induced by dust in the halo of $\sim L_\star$  galaxies at the mean redshift of our sample.
  The three blue curves represent extrapolations to higher redshifts, using
  different dust models: a constant co-moving density (dashed line), a density
  decreasing with the mean metallicity (solid light blue) and a suppressed density
  at high redshift (dark blue). Various observational constraints are shown and described
  in Appendix~B.}
\label{fig:cosmic_Av}
\end{center}
\end{figure}
%-----------------------------------------------------------------------------
Assuming that most of the dust in the Universe is associated with
galaxies and that most of the optical depth for extinction originates
from the halo of galaxies with $L\sim L_{\rm eff}$ (defined in
Equation~\ref{eq:def_L}), the mean dust optical depth in the Universe
up to a redshift $z$ is then given by
\begin{eqnarray}
\bar\tau(\lambda,z)&=&  \int^z_0\,
\sigma\,n\,
\bar\tau_{\rm g}\left(\frac{\lambda}{1+z}\right)\,\frac{c\;(1+z)^2}{H(z)}\, {\rm d} z
\label{eq:tau}
\end{eqnarray}
where $\sigma$ and $n$ are the cross-section and number density
of galaxies with $L\sim L_{\rm eff}$.
For a dust profile extending up to the virial radius, $r_v$, of
these galaxies we have $\sigma=\pi\,r_v^2$ and
$\bar\tau_g$ is the average optical depth within the halo, which
we define with $20\,h^{-1}\,{\rm kpc}<r<r_v$.
We evaluate $\bar\tau(\lambda,z)$ (Equation~\ref{eq:tau}) and present
the results in terms of extinction in the observer-frame $V$-band.  
Integrating Equation \ref{eq:tau} up to the mean redshift of the galaxy sample
used in this analysis, we find the value denoted by the blue circle:
$\langle{\rm A_V}(z=0.36) \rangle\simeq 0.02\;{\rm mag}$.
We then estimate ${\rm A_V}(z)$ for different dust models shown in Figure~\ref{fig:cosmic_Av}:
\begin{enumerate}
\item we first use a constant dust density with redshift, represented with the dashed line.
At high redshift, such an estimate is expected to represent an upper limit
on the allowed extinction.
\item A more realistic estimate should take into account the fact that the
amount of dust in and around galaxies is redshift dependent as dust is
being produced at a rate following that of the metals.  As mentioned
above, useful constraints on the evolution of the amount of dust
around galaxies come from studies of metal absorbers.
Recently, \cite{2008MNRAS.385.1053M} probed the redshift
evolution of the amount of dust associated with MgII absorbers
from $z=0.4$ to $z=2$. They found that their dust content follows
$\rho_{\rm dust}\propto (1+z)^{-1.1}$, which turns out to be similar
to the evolution of cosmic star density.  By taking this redshift
dependence into account, we obtain a 
an alternative estimate of the cosmic opacity. 
The corresponding results are shown in
Figure~\ref{fig:cosmic_Av} with the solid light-blue curve. As can be seen,
taking this effect into account brings a modest change to our previous
estimate and lowers the total opacity by about a factor two at
$z\sim2$. 
\item Finally, to illustrate the range of possible redshift dependences,  we consider a third model where we damp the density of dust by an addition factor $(1+z)^{-1}$, shown with the dark blue curve. This model is somewhat \emph{ad hoc} in nature, but demonstrates that the current measurements and limits allow for considerable variation at higher redshift.
\end{enumerate}
Figure~\ref{fig:cosmic_Av} also shows various upper and lower limits (detailled in Appendix B) on the
opacity as a function of redshift.  As can be seen, at $z\sim1$, $A_V$ values are bounded within about a factor $10$.

\subsubsection{Implications for supernova experiments}

Type Ia supernovae provide us with an estimate of luminosity distances and 
have been extensively used to constrain cosmological parameters, dark energy in particular. 
Supernovae are standardizable candles and become usable as standard candles after re-normalizing their brightness for intrinsic brighter-bluer and brighter-slower trends as well
as dust extinction. We now investigate how the presence of intergalactic dust can affect such constraints.

The color $c$ of each supernovae is the sum of several contributions: 
$c=\sum_i c_i$, where $c_i$ are the intrinsic color, dust reddening by the host, 
dust along the line-of-sight, etc. Each of them can be corrected for using an appropriate
"reddening-to-extinction" coefficient $\beta_i$. The observed supernova magnitudes are used as a distance estimator according to
\begin{equation}
\mu_i=m_i-M+\alpha(s_i-1)-\beta\,c_i
\label{eq:sn}
\end{equation}
where the apparent magnitude $m_i$, the stretch $s_i$ and the color $c_i$ are derived from the fit to the light curves. The parameters $\alpha$, $\beta$ and the absolute magnitude $M$ are fitted by minimizing the residuals in the Hubble diagram.
A color excess $c_i$ which does not contribute to a significant scatter will not affect the inferred value of $\beta$. Its reddening-to-extinction coefficient will be described by the best-fit $\beta_0$ and might lead to a bias if its intrinsic $\beta$ differs from this value.
As shown above, the presence of intergalactic dust might introduce such an effect.
On large scales around galaxies we find $R_V\sim4$, which corresponds to a value of $\beta\sim5$, while SN analyses lead to $\beta\sim2-3$ \citep{2007ApJ...664L..13C,2008ApJ...686..749K}.
This indicates that, at some level, the (redshift-dependent) contribution of intergalactic might not be properly corrected by using Eq. \ref{eq:sn}. The magnitude of this effect and its impact on dark energy constraints must be investigated and quantified.

%==================================================
\subsection{Dust-to-light ratio}
%==================================================

\begin{figure}
\begin{center}
  \includegraphics[width=.9\hsize]{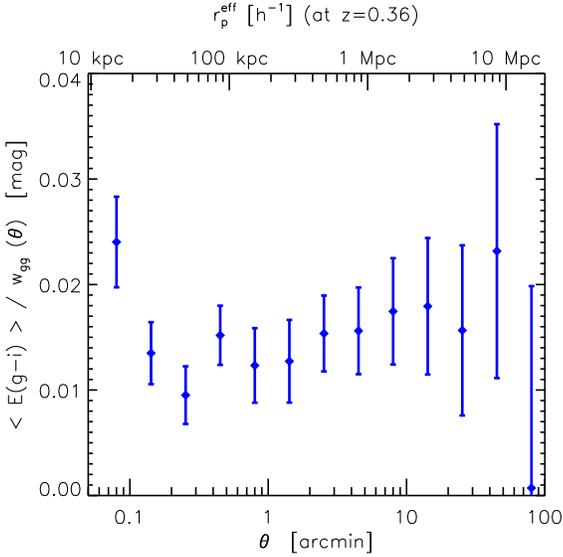}
  \caption{Ratio between the reddening-galaxy correlation and the
galaxy autocorrelation function, indicating that the dust to light
ratio is weakly scale dependent.}
\label{fig:color_bias}
\end{center}
\end{figure}

We can compare the statistical properties of the light (traced by
galaxies) and dust distributions by computing the ratio of the
galaxy-reddening cross-correlation to the autocorrelation of the same
galaxies.  To do so we compute, as a function of scale, the parameter
\begin{eqnarray}
	\beta_{(g-i)} (\theta)
	&=&\frac{\langle E_{gi}\rangle (\theta)}
	{w_{gg}(\theta)}
\end{eqnarray}
where we use the $g-i$ color as our estimate of dust reddening (see
section \ref{sec:profile}). This quantity is shown in
Figure~\ref{fig:color_bias}.  The dust reddening-to-light ratio
appears to be weakly scale-dependent.  On scales greater than about
one arcminute, we find
\begin{equation}
\beta_{(g-i)} (\theta>1') \simeq 0.015~\mathrm{mag}\;.
\end{equation}
The excess seen on the smallest scale might be due to a contribution
from galactic disks.  The dust-to-light ratio is a quantity which can
be used to constrain models describing the transport of dust outside
of galaxies.  Our results show that, on large scales, the dust distribution 
follows that of galaxies.
%-----------------------------------------------------------------------------
%-----------------------------------------------------------------------------

%==================================================
%==================================================
\section{SUMMARY}
%==================================================
%==================================================

Intervening galaxy halos and large-scale structure affect the light of 
background sources through both gravitational lensing and dust extinction.
In this paper we present simultaneous detections of these two effects
obtained by measuring the cross-correlation between the 
brightness and colors of about 85,000 $z>1$ quasars and 20 million 
$z\sim 0.3$ galaxies from the SDSS.  We find that quasar brightness is 
correlated with galaxy overdensity, indicating that magnification effects 
dominate over dust extinction for this sample of galaxies.  
By examining the quasar colors, we observe that quasars appear to be redder
as the projected galaxy density increases.  Both effects are detected on
scales ranging from 0.1 arcmin to about 2 degrees, corresponding to
projected radii of 20\,{\rm $h^{-1}\,$kpc} to $20\,h^{-1}{\rm Mpc}$ at the mean redshift of
the galaxy sample.  More specifically, we find that
\begin{itemize}
\item galaxies and large-scale structures at $z\sim0.3$ induce an excess
reddening to background sources given by
$\langle \rm{E(B-V)}\rangle \simeq 1.4\times10^{-3}\,\left({\theta}/{1\arcmin}\right)^{-0.84}$.
On arcminute-scales around galaxies, the slope of the extinction curve is 
constrained by ${\rm R_V}=3.9\pm2.6$, a value consistent with that of our
Galaxy. Assuming the Galactic value of
${\rm R_V}=3.1$, we find a mean extinction profile given by
$\langle {\rm A_V}\rangle \simeq 2.4\times10^{-3}\,\left({r_p}/{100\,h^{-1}\,{\rm kpc}}\right)^{-0.84}$.
The ratio of the galaxy-reddening to galaxy-galaxy correlation functions
is found to be roughly scale independent, with an amplitude
$\langle {\rm E}(g-i) \rangle/w_{gg}\simeq 0.015$ mag.
\item The amplitude of dust extinction in the $V$-band is about one-third of that
of the \emph{observed} brightening due to magnification.  We estimate the
extinction-corrected magnification profile and find
$\langle \mu(\theta) \rangle \simeq 0.025\,(\theta/{\rm 1 \arcmin})^{-0.8}$, 
consistent with the results of \cite{2005ApJ...633..589S} and extending those 
to both smaller and larger scales. The average mass surface density 
profile of the galaxies inferred from our measurements is comparable to 
galaxy-galaxy lensing estimates.
\item At a separation of $20$ $h^{-1}\,$kpc from a $z\sim0.3$ galaxy, a background
source is, on average, magnified by a factor $\mu=1.15$ and reddened by
E(B-V)$\simeq 0.01$.
\end{itemize}

The detection of dust on large-scales has a number of implications:
\begin{itemize}
\item The amount of dust found in galactic halos is found to be
  comparable to that in the disk of $\sim L^\star$ galaxies and
  therefore significantly larger than that of dwarf satellites. This
  implies the existence of a diffuse component of dust in halos,
  predicted by some models of dust halo dynamics but heretofore
  unobserved.
\item Having argued that the dominant contribution of dust reddening
  observed in this analysis is due to galaxies with $L\sim
  0.5\,L^\star$, we have estimated their halo dust mass to be about
  $5\times 10^7\,M_\odot$ within their virial radius.
\item Including both disk and halo contributions, we find $\Omega_{\rm
    dust}\simeq 5 \times10^{-6}$, a value roughly twice that estimated
  by \citep{2004ApJ...616..643F} for galactic disks.
\item Such an extended distribution of dust around galaxies will
affect the apparent magnitude of distant sources. 
We have estimated the mean opacity of the Universe due to dust in 
galactic halos. Our model-dependent estimate gives
${\rm A_V}(z=0.5)\sim 0.03$ mag. Such a value is less constrained
at higher redshifts due to our limited knowledge of the evolution of dust
with redshift. Considering several models, we found
${\rm A_V}(z=1)\sim 0.05-0.09$ mag.
This will affect the brightness estimates of Type Ia supernovae at
high redshift, which require high precision in order to maximize their
constraints on cosmological parameters.  Dust reddening may also induce
dispersion in brightness and color.
\end{itemize}

The technique presented in this paper provides us with a unique probe 
of the distribution of dust (warm and cold) on large scales around galaxies, which is otherwise
difficult to explore. It opens up the way to studies of the amount of circumgalactic dust as a function of galaxy type, luminosity and environment which may shed light on the 
origin of the dust. Extending the analysis to UV measurements is an important
task to obtain better constraints on the dust properties. 
The final remark is that this analysis requires only accurate photometric data 
in several passbands.

%==================================================
%==================================================
\section*{Acknowledgements}

We thank Robert Lupton, Jim Gunn, Joseph Weingartner, Bruce Draine,
Doron Chelouche, Tony Tyson, Erin Sheldon, and
Latham Boyle for useful discussions. MF is supported by the Monell
Fundation and the Friends of the Institute for Advanced Study.

Funding for the SDSS and SDSS-II has been provided by the Alfred
P. Sloan Foundation, the Participating Institutions, the National
Science Foundation, the U.S. Department of Energy, the National
Aeronautics and Space Administration, the Japanese Monbukagakusho, the
Max Planck Society, and the Higher Education Funding Council for
England. The SDSS Web Site is http://www.sdss.org/.

%==================================================
%==================================================
\begin{appendix}

\section{Derivation of $C_S$}
\label{appendix_A}

The observable magnitude shift defined by Equation~\ref{eq:obs} is a
function of the shape of the magnitude distribution, the limiting
magnitude and the induced magnitude shift $\delta m$.  In the case
where the induced magnitude shift $\delta m$ is small compared to the
limiting magnitude of the sample, the difference between the observed
and induced magnitude shift can be linearized in $\delta m$. We have
\begin{eqnarray}
\Delta m_\mathrm{obs}
&=&\left\langle{m}\right\rangle-\left\langle{m_0}\right\rangle \nonumber\\
&=&
\frac{ \int \d m\, m\, n(m-\delta m) }{ \int \d m\, n(m-\delta m) }
-\frac{ \int \d m\, m\, n(m) }{ \int \d m\, n(m) }
\end{eqnarray}
For induced magnitude shifts small compared to unity, we can Taylor-expand the
above expression to first-order in $\delta m$:
\begin{eqnarray}
\Delta m_\mathrm{obs}
&\simeq& \frac{\delta m}{I_0}\,
\left[  \frac{I_1}{I_0} \, \int \d m\, n'(m) - \int \d m\,n'(m)\,m
\right ]\,
\end{eqnarray}
where $I_\alpha=\int \d m\, n(m)\,m^\alpha$. The above expression can be simply
written as
\begin{eqnarray}
\Delta{m_\mathrm{obs}}&\simeq&\delta m\times
\mathrm{C_S}\,.  
\end{eqnarray}
with
\begin{equation}
{\rm C_S}= 1-
\frac{1}{\rm N_0^{tot}} \, \frac{\d {\rm N}}{\d m}(m_\ell)\,
\big[m_\ell-\langle m_0 \rangle \big]
\end{equation}
We can verify that in the case of a power-law luminosity function,
i.e. $dN/df \propto f^\alpha$ or $dN/dm \propto a^m$, the above
expression gives $C_S=0$ which implies that no magnitude shift can be
observed.

%==================================================
\section{Observational constraints on the cosmic transparency}
\label{appendix_B}

In this appendix, we detail the values of the cosmic opacity upper and lower 
limits used in Figure~\ref{fig:cosmic_Av}.  Note that the upper limit values 
plotted in the figure reflect the 95\% or 99\% confidence limits for those
measurements, while the lower limit values are taken from the actual measured
values.
\begin{itemize}

\item \cite{2008arXiv0810.5553M} put a virtually assumption-free constraint on 
the opacity of the Universe at low redshift using the Tolman test:
\begin{equation}
D_L = (1+z)^2\;D_A,
\end{equation}
where $D_L$ is the luminosity distance and $D_A$ the angular diameter distance, 
independent of world model.  Any observed deviation from this expected 
relation is taken to be a result of extinction along the line of sight.
Combining observational results from supernovae and baryon acoustic 
oscillations to estimate the change in optical depth from redshift 0.20 to 
0.35, they found that $\Delta \tau<0.13$ at 95\% confidence.
Assuming no evolution of the dust properties in the redshift range $0<z<0.25$, 
we can use their result to estimate the expected $A_V$ extinction up to 
$z\sim0.35$:
\begin{eqnarray}
A_V\simeq 1.08\;\Delta \tau \times
\left [1+ {\rm I}_{0}^{0.2}/{\rm I}_{0.2}^{0.35}
\right ]
\end{eqnarray}
where ${\rm I}_x^y=\int_{x}^{y} {\rm d}z\;{(1+z)^2}/{H(z)}$. 
This gives $A_V(z=0.35)<0.14$ mag.

\item \cite{2009arXiv0902.2006A} obtained a constraint on the low-redshift 
opacity by combining observational results from supernovae and the Hubble Key 
project.  They simultaneously fitted for $H(z)$ and the luminosity distance,
where the latter quantity was allowed to be modulated extinction. Using the 
Union sample of supernovae \citep{2008ApJ...686..749K} with a mean redshift of 
$z \sim 0.5$, they found
\begin{eqnarray}
A_V \simeq 1.08\times\,2\,\epsilon\,z
\end{eqnarray}
with $\epsilon=-0.01^{+0.08}_{-0.09}$.  This gives $A_V(z\sim0.5)<0.08$ mag. at
99\% confidence.

\item \cite{2003JCAP...09..009M} analyzed the scatter in quasar colors as a 
function of redshift, attributing its excess to dust extinction.  They 
reported a $99\%$ upper limit on the cosmic opacity: $A_V(z=1)<0.2$ mag.

\item \cite{2008MNRAS.385.1053M} quantified the mean amount of reddening and 
extinction induced by strong MgII absorbers, i.e. systems usually found within 
$\sim 100$ kpc of $\sim L^\star$ galaxies \citep{2007ApJ...658..161Z}. They were 
able to constrain the mean reddening $\langle E_{B-V} \rangle$ as a function of 
the rest equivalent width $W_0$ of the absorbers and their redshift, in the 
range $0.4<z<2$.  In addition, as described in section \ref{sec:implications}, 
various authors have shown that, on average, the extinction curve associated 
with MgII absorbers is consistent with that of the SMC, i.e. with $R_V\simeq3$.
The mean extinction induced by these systems can then be computed according to
\begin{equation}
A_V(z) = R_V\;\int_0^\infty \d W_0 \int_0^z \d z \;\frac{\d^2 N}{\d W_0 \d z}\;\langle E_{B-V}(W_0,z) \rangle\,.
\end{equation}
Using the parameterization of the the incidence rate of MgII absorbers, 
$\d^2N/\d W_0\d z$, given by \cite{2005ApJ...628..637N}, and assuming no 
evolution in the dust properties in the range $0<z<0.4$, 
we can use these measured values as lower limits on the global extinction.
We find: ${\rm A_V}(z=0.6)>0.009$ mag, ${\rm A_V}(z=1.3)>0.029$ mag and 
${\rm A_V}(z=1.8)>0.044$ mag.  

\end{itemize}

\end{appendix}

%==================================================
%==================================================
%==================================================
%==================================================
%==================================================

\end{document}